\documentclass[journal]{IEEEtran}

\usepackage{cite}
\usepackage{amsmath,amssymb,amsfonts}
\usepackage{algorithmic}
\usepackage{graphicx}
\usepackage{textcomp}
\usepackage{xcolor}
\usepackage{url}
\usepackage{subcaption}
\usepackage{booktabs}
\usepackage{multirow}
\usepackage{makecell}
\usepackage{threeparttable}
\usepackage{stfloats}
\usepackage{hyperref}

\title{Distributed Multi-View Vision-Only RSSI Estimation}

\author{Jung-Beom Kim and Woongsup Lee%
\thanks{Jung-Beom Kim is with the Graduate School of Information, Yonsei University, Seoul 03722, Republic of Korea (e-mail: kjung99@yonsei.ac.kr).}%
\thanks{Woongsup Lee is with the Graduate School of Information, Yonsei University, Seoul 03722, Republic of Korea (e-mail: woongsup.lee@yonsei.ac.kr).}}

\begin{document}

\maketitle

\begin{abstract}
Received Signal Strength Indicator (RSSI) estimation is essential for wireless link management, yet conventional feedback-based approaches incur uplink overhead, suffer from measurement instability, and are subject to inherent feedback loop latency, rendering proactive adaptation infeasible. Although vision-based approaches have been explored, existing methods remain limited by hardware dependency or auxiliary inputs, and lack the spatial diversity needed to resolve camera-side NLoS conditions. To address these limitations, we propose MulViT-TF, a vision-only RSSI estimation framework that exploits distributed multi-view observations through Transformer-based fusion, achieving complementary spatial coverage without any auxiliary sensing inputs. Experimental results across two distinct indoor scenes demonstrate that MulViT-TF achieves RMSE reductions of up to 26.3\% and improves the 3dB error coverage by up to 13.8 percentage points over the best-performing single-view baseline, while using fewer FLOPs and parameters.
\end{abstract}

\begin{IEEEkeywords}
Received Signal Strength Indicator, Distributed Multi-View Fusion, Indoor Wireless Sensing, Vision Transformer, Out-of-Band Information, WiFi
\end{IEEEkeywords}

\section{Introduction}
\IEEEPARstart{R}{eceived} Signal Strength Indicator (RSSI) is a critical metric in wireless communication, playing a key role in various operations, including handover~\cite{zhang2015qoe}, beam management~\cite{zhang2023beam}, and transmit power control for energy efficiency~\cite{moreno2024grey}. Conventionally, RSSI is obtained through receiver-side feedback; however, this approach consumes uplink resources, suffers from measurement instability due to multipath fading and shadowing, and inherently reports past channel conditions due to feedback loop latency, precluding proactive adaptation. Inspired by the concept of out-of-band information aided wireless communications~\cite{gonzalez2017mmwave}, visual environmental observations have emerged as a particularly promising source of spatial context~\cite{kim2024role}. Since wireless signal propagation is fundamentally shaped by the spatial configuration of the environment, including the positions of obstacles and persons, camera images provide a direct and passive means of capturing such information eliminating the need for receiver-side feedback. Leveraging visual environmental observations has thus shown promise not only for RSSI estimation and prediction~\cite{yan2025reading,nguyen2025distributed,Xzhang2025vision, nishio2019proactive} but also for a broader range of wireless tasks, including handover~\cite{charan2021vision}, channel estimation and prediction~\cite{Qzhang2025vision,kim2026vision,shi2025weicsip}, beamforming~\cite{ahn2022toward,xu2022computer,li2026outofband}, and codebook design~\cite{chen2022codebook}.

Existing studies on wireless communication leveraging visual information commonly suffer from several limitations. First, several approaches require additional inputs beyond camera images — such as depth images~\cite{nishio2019proactive, kim2026vision, ahn2022toward}, LiDAR~\cite{nguyen2025distributed}, GPS coordinates~\cite{yan2025reading, li2026outofband}, or explicit feedback signals~\cite{nguyen2025distributed, charan2021vision, shi2025weicsip} — which either limit deployability due to specialized hardware requirements or introduce continuous communication overhead and synchronization burden during inference. Second, prior works are either limited to single-view settings, which suffer from restricted Field-of-View (FoV) coverage and Non-Line-of-Sight (NLoS) conditions~\cite{yan2025reading, Xzhang2025vision, nishio2019proactive, Qzhang2025vision, ahn2022toward}; mobile single-view approaches can partially mitigate these limitations, but still depend on a single instantaneous viewpoint during inference, leaving FoV and NLoS constraints fundamentally unresolved~\cite{nguyen2025distributed, chen2022codebook}. Furthermore, multi-view configurations partially address FoV limitations but still fail to resolve NLoS conditions, as centralized deployments lack sufficient spatial diversity~\cite{kim2026vision, shi2025weicsip, xu2022computer}; distributed deployments process each view independently without cross-view fusion, failing to effectively exploit spatially distributed cameras and thus reducing to multiple independent single-view systems~\cite{charan2021vision, li2026outofband}. Real-world indoor environments are increasingly equipped with distributed RGB camera-based IoT devices, such as CCTV systems and service robots, which naturally provide multi-view observations. Building upon this, leveraging distributed multi-view observations for wireless communication enables spatially diverse and complementary observations across NLoS regions that are otherwise restricted by limited FoV coverage and Line-of-Sight (LoS) dependency.

In this letter, we propose MulViT-TF, a vision-only RSSI estimation framework that leverages spatially distributed RGB cameras, without requiring specialized hardware or auxiliary signal measurements. Independent ViT encoders process per-camera observations, and a Transformer-based fusion module captures cross-camera spatial correlations, enabling robust estimation by leveraging complementary viewpoints across spatially distributed cameras. The proposed framework is validated through real-world experiments using actual RGB camera and WiFi RSSI data collected across two distinct indoor scenes. Compared to the best-performing single-view baseline, MulViT-TF reduces RMSE by 26.3\% and 16.6\% and raises the proportion of estimates within a 3dB error bound from 70.7\% to 84.5\% and from 70.9\% to 78.4\% in Scene1 and Scene2, respectively, despite using fewer FLOPs and parameters. The source code and pretrained models are publicly available at \href{https://github.com/Kim-JungBeom/distributed-multi-view-rssi-estimation}{https://github.com/Kim-JungBeom/distributed-multi-view-rssi-estimation}, and the hardware implementation and time synchronization pipeline (Raspberry Pi and ESP32) at \href{https://github.com/Kim-JungBeom/esp32-csi-raspi-cam-sync-collector}{https://github.com/Kim-JungBeom/esp32-csi-raspi-cam-sync-collector}.

The remainder of this letter is organized as follows. Section~\ref{sec:proposed} presents the proposed MulViT-TF framework. Section~\ref{sec:experiments} provides the experimental results and analysis. Finally, Section~\ref{sec:conclusion} concludes the letter.

\section{Proposed Framework}
\label{sec:proposed}
\begin{figure*}[t]
    \centering
    \includegraphics[width=0.95\textwidth]{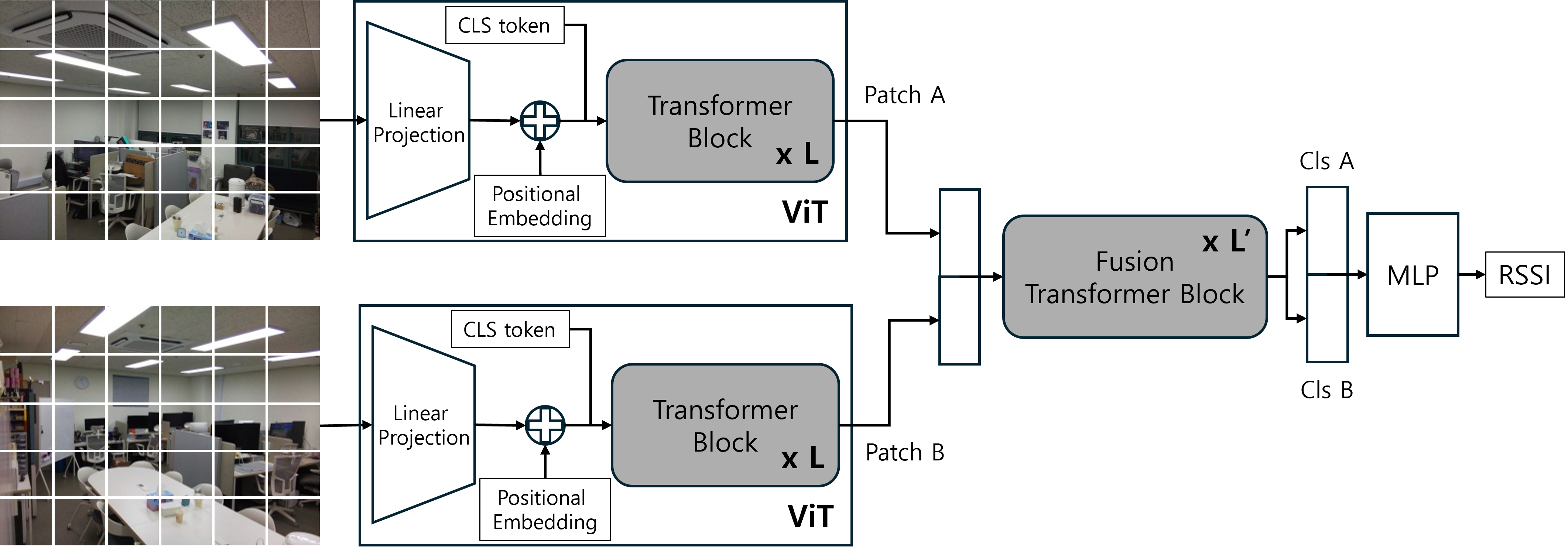}
    \caption{Overall architecture of MulViT-TF with distributed multi-view ViT and Transformer-based fusion.}
    \label{fig:ViT}
\end{figure*}
To leverage spatial information from multiple distributed viewpoints, we propose a multi-view framework for RSSI estimation in which per-camera ViT~\cite{dosovitskiy2020image} encoders extract viewpoint-specific features that are subsequently fused by a Transformer~\cite{vaswani2017attention} module. Fig.~\ref{fig:ViT} illustrates the two-view case of MulViT-TF as an example, showing two independent ViT encoders and a Transformer-based fusion module. Each camera image is divided into $N$ patches and embedded into a $D$-dimensional space. A learnable CLS token $x^{cls}$ is prepended to the patch sequence to serve as a global representation of the input image, and positional embeddings are added as follows:
\begin{equation}
\mathbf{z}_0 = [x^{cls}; x^1_p E; x^2_p E; \cdots; x^N_p E] + E_{pos},
\end{equation}
where $x^i_p \in \mathbb{R}^{P^2 \cdot C}$ is the $i$-th patch, $P$ is the patch size, $C$ is the number of input channels, $E \in \mathbb{R}^{P^2C \times D}$ is the linear projection matrix, and 
$E_{pos}$ is the positional embedding. The resulting sequence $\mathbf{z}_0$ is passed through $L$ Transformer blocks, where each block applies Multi-Head Attention (MHA) and a Feed-Forward Network (FFN) as follows:
\begin{equation}
\mathbf{z}'_l = \mathrm{MHA}(\mathrm{LN}(\mathbf{z}_{l-1})) + \mathbf{z}_{l-1},
\end{equation}
\begin{equation}
\mathbf{z}_l = \mathrm{FFN}(\mathrm{LN}(\mathbf{z}'_l)) + \mathbf{z}'_l,
\end{equation}
where $\mathrm{LN}(\cdot)$ denotes Layer Normalization (LN). The final output $\mathbf{z}_L \in \mathbb{R}^{(N+1) \times D}$ is the token matrix after $L$ Transformer blocks. The token matrices from both cameras are then concatenated to form the fusion input:
\begin{equation}
\mathbf{Z} = \begin{bmatrix} \mathbf{z}^A \\ \mathbf{z}^B \end{bmatrix} 
\in \mathbb{R}^{2(N+1) \times D},
\end{equation}
where $\mathbf{z}^A, \mathbf{z}^B \in \mathbb{R}^{(N+1) \times D}$ are the token matrices of cam A and cam B, respectively. The concatenated token matrix $\mathbf{Z}$ is then processed by the fusion Transformer block, which adopts the same number of attention heads and embedding dimension $D$ as the individual ViT encoders, but takes $M(N+1)$ patches as input, scaling proportionally with the number of cameras $M$. The fusion block is stacked $L'$ times, yielding output $\mathbf{Z}'$. By allowing each token to attend to all tokens from both cameras, the MHA enables the model to capture cross-camera spatial correlations that cannot be exploited by simple feature concatenation. The CLS token vectors are then extracted from $\mathbf{Z}'$ 
for each camera, $\mathbf{v}^A_{cls} = \mathbf{Z}'[0] \in \mathbb{R}^{D}$ and $\mathbf{v}^B_{cls} = \mathbf{Z}'[N+1] \in \mathbb{R}^{D}$, and concatenated to form the fused feature vector:
\begin{equation}
\mathbf{f} = [\mathbf{v}^A_{cls}; \mathbf{v}^B_{cls}] \in \mathbb{R}^{2D},
\end{equation}
which is passed through a Multi-Layer Perceptron (MLP) head to produce the final RSSI estimate $\hat{y} = \mathrm{MLP}(\mathbf{f})$.

Each ViT encoder is initialized from the same checkpoint pre-trained on the Places365 dataset~\cite{zhou2017places} using an image classification task following the DeiT training recipe~\cite{touvron2021training}, and subsequently fine-tuned independently with unshared weights to allow each encoder to specialize to its respective camera viewpoint. Places365 is chosen because it is a scene-centric dataset that requires holistic understanding of the entire environment rather than object-level recognition, which aligns well with our indoor RSSI estimation task where the overall spatial configuration governs signal propagation. The DeiT training recipe is adopted for its data-efficient training strategy, enabling effective ViT optimization without large-scale data. This pre-training enables the model to learn general visual representations of indoor environments prior to fine-tuning. The pre-trained backbone is then fine-tuned on the target RSSI estimation task in two phases. In the first phase, the backbone is frozen and only the CLS token, positional embeddings, fusion module, and MLP head are trained. In the second phase, the entire network is jointly optimized in an end-to-end manner, with a reduced learning rate applied to the backbone to preserve the pre-trained representations. In both phases, a cosine annealing learning rate schedule is applied~\cite{loshchilov2017sgdr}, and the model is trained by minimizing the Mean Squared Error (MSE) loss:
\begin{equation}
    \mathcal{L}_{\text{MSE}} = \frac{1}{N} \sum_{i=1}^{N} \left( \hat{y}_i - y_i \right)^2
\end{equation}
where $N$ is the number of training samples, and $\hat{y}_i$ and $y_i$ are the estimated and ground-truth RSSI of the $i$-th sample, respectively.

To evaluate the effectiveness of the proposed Transformer-based fusion, we consider two baseline models. The first baseline is a single-camera model, which employs a single ViT encoder and directly passes the CLS token vector through an MLP head to estimate the RSSI value, without any cross-camera interaction. The second baseline, MulViT-TWDNN, replaces the fusion Transformer block with a token-wise DNN. A learnable segment embedding is first added to each camera's token sequence to distinguish camera identity, and the resulting token matrices are concatenated to form $\mathbf{Z} \in \mathbb{R}^{2(N+1) \times D}$, which is then processed by the token-wise DNN consisting of stacked residual blocks — each comprising LN and an FFN — applied identically across all token positions. The CLS tokens are then extracted and passed through an MLP head for RSSI estimation.
\section{Experimental Results}
\label{sec:experiments}
\begin{figure*}[t]
    \centering

    \begin{subfigure}[t]{0.48\textwidth}
        \centering
        \includegraphics[width=\linewidth]{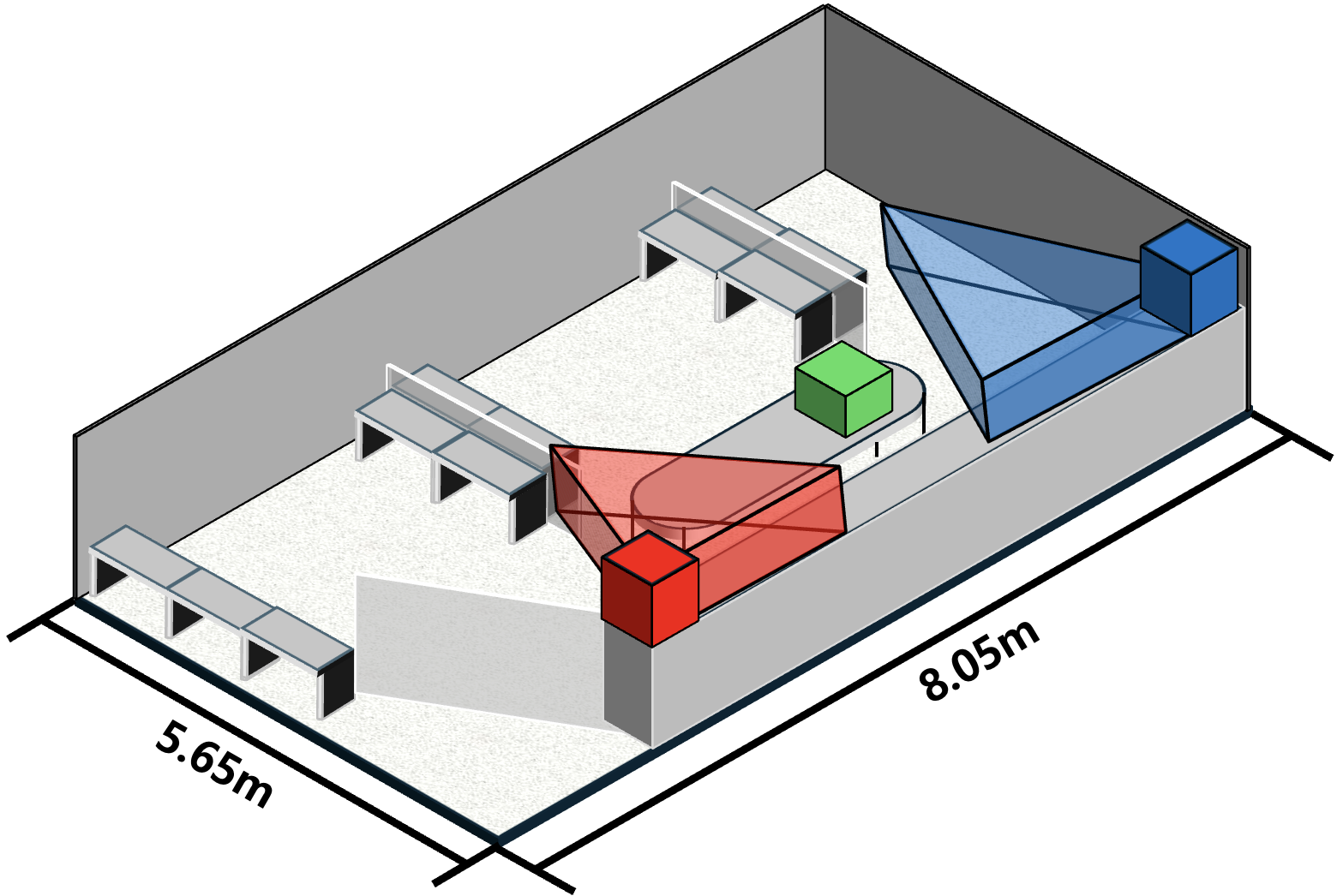}
        \caption{Scene 1 Layout}
    \end{subfigure}
    \hfill
    \begin{subfigure}[t]{0.48\textwidth}
        \centering
        \includegraphics[width=\linewidth]{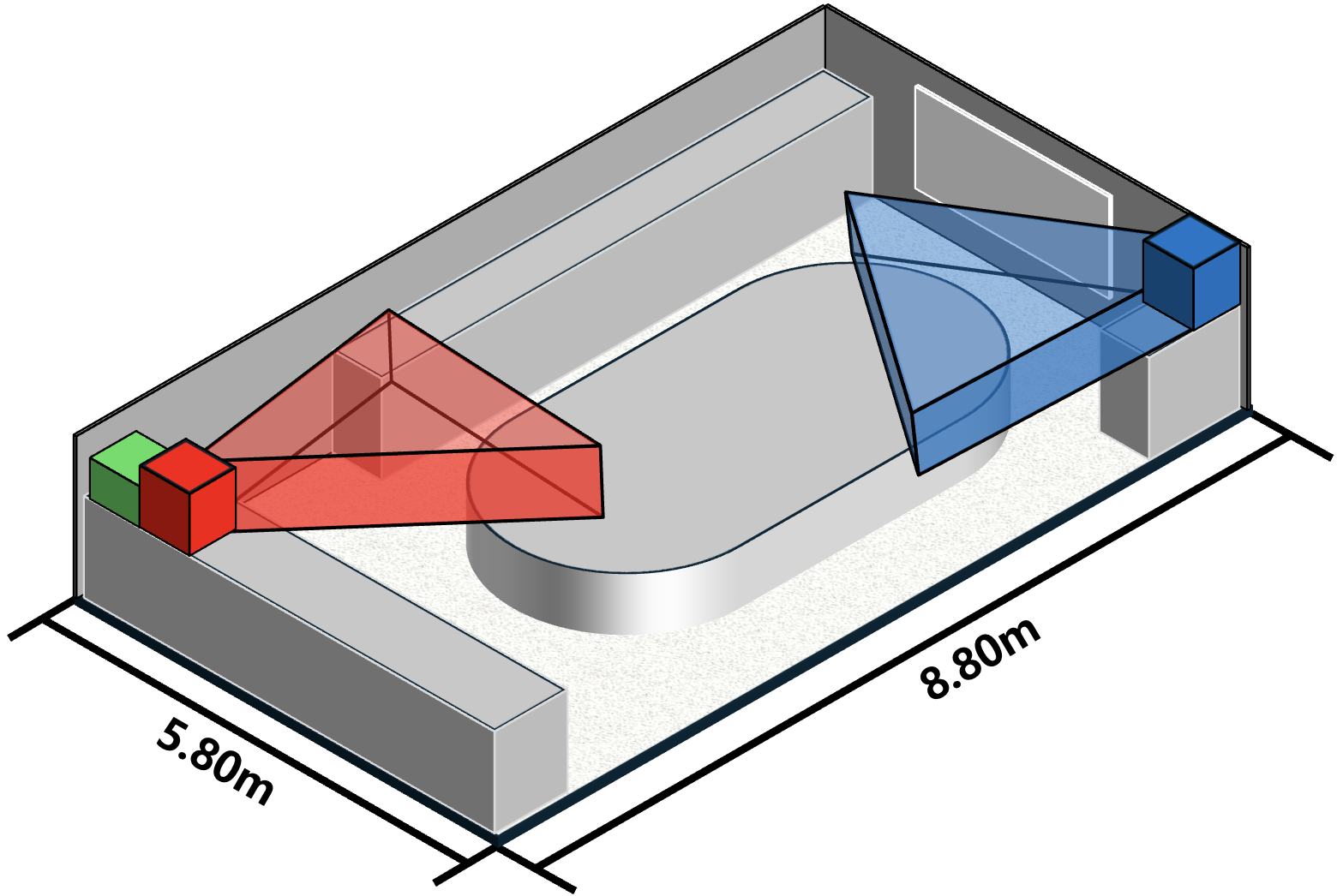}
        \caption{Scene 2 Layout}
    \end{subfigure}

    \vspace{0.5em}

    \begin{minipage}[t]{0.49\textwidth}
        \centering
        \begin{subfigure}[t]{0.49\textwidth}
            \includegraphics[width=\textwidth]{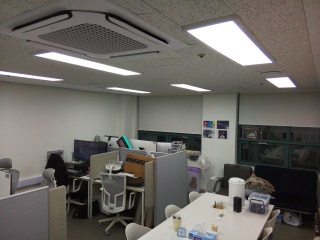}
            \caption{Scene 1, CAM1}
        \end{subfigure}
        \hfill
        \begin{subfigure}[t]{0.49\textwidth}
            \includegraphics[width=\textwidth]{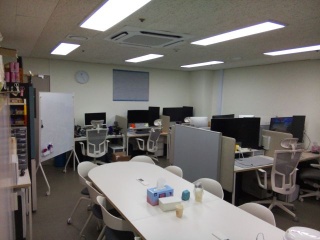}
            \caption{Scene 1, CAM2}
        \end{subfigure}
    \end{minipage}
    \hfill
    \begin{minipage}[t]{0.49\textwidth}
        \centering
        \begin{subfigure}[t]{0.49\textwidth}
            \includegraphics[width=\textwidth]{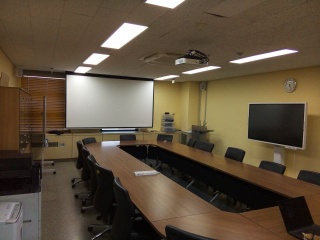}
            \caption{Scene 2, CAM1}
        \end{subfigure}
        \hfill
        \begin{subfigure}[t]{0.49\textwidth}
            \includegraphics[width=\textwidth]{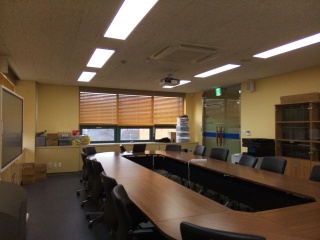}
            \caption{Scene 2, CAM2}
        \end{subfigure}
    \end{minipage}

    \caption{Indoor measurement environments and corresponding distributed camera viewpoints for Scene 1 and Scene 2, where red, blue, and green blocks denote CAM1, CAM2, and AP, respectively, and colored regions indicate the FoV of each camera.}
    \label{fig:scenes}
\end{figure*}

\begin{figure}[t]
    \centering
    \includegraphics[width=\columnwidth]{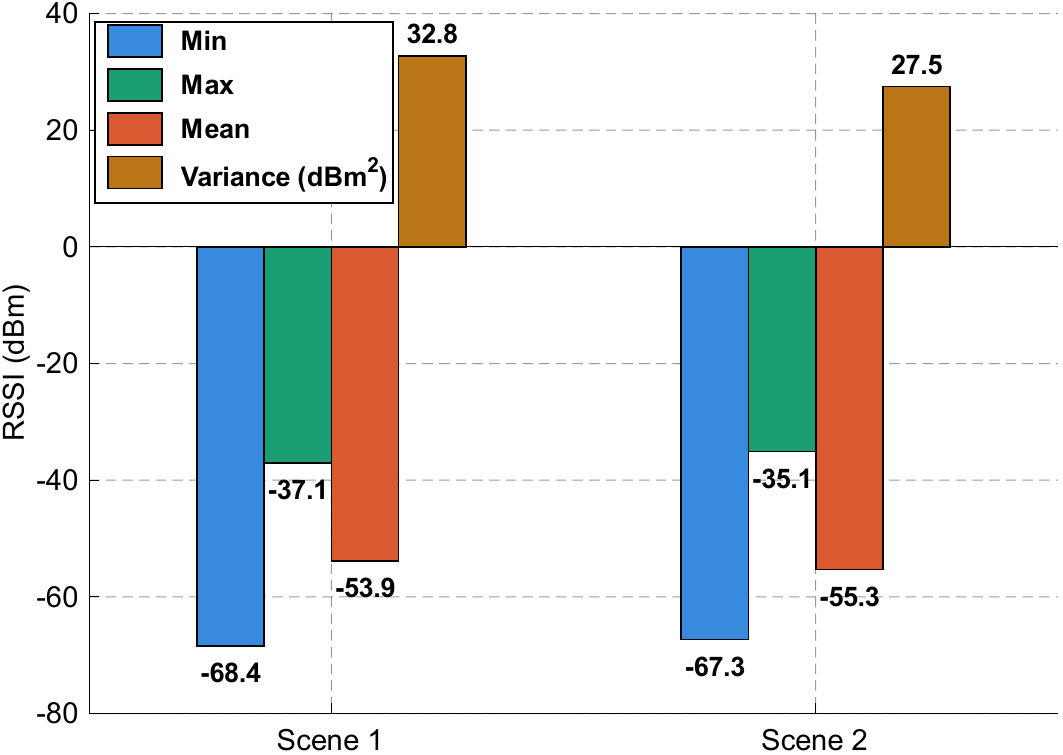}
    \caption{RSSI statistics across different indoor environments.}
    \label{fig:rssi_stats}
\end{figure}

The experiments were conducted in two distinct indoor environments, as shown in Fig.~\ref{fig:scenes}. Scene 1 is an office room where the Access Point (AP) placement results in a higher proportion of NLoS conditions between the AP and Station (STA) due to furniture obstacles. Scene 2 is a conference room characterized by predominantly LoS conditions between the AP and STA. Fig.~\ref{fig:rssi_stats} shows the RSSI statistics across the two experimental scenes. In both scenes, a single STA moves throughout the space while the distributed cameras capture the surrounding indoor environment including the location of the STA-carrying user. We assume a single-user scenario where the target STA is known in advance through the AP-STA link association. The hardware setup consists of two Raspberry Pi Camera Module V2 and two ESP32-S3 FeatherS3 modules serving as the AP and STA, each equipped with a 2.4GHz omnidirectional dipole antenna (2.1dBi).

Camera images were captured at 20Hz, yielding 10,000 samples per scene. Input images were resized to $320\times240$ to preserve the full FoV. RSSI values were obtained via the AP-STA link at 40 Hz, resulting in approximately 20,000 raw samples per scene. The raw RSSI samples were preprocessed through: (i) linear interpolation using the nearest neighboring valid samples to recover missing values, (ii) median absolute deviation-based outlier removal with a local window of 40 samples (1s) and a threshold of 5, and (iii) Gaussian smoothing with 4 samples (100ms) to suppress instantaneous fluctuations caused by measurement noise, following a similar pipeline to ~\cite{nguyen2025distributed}. The preprocessed RSSI samples were downsampled to 20Hz by averaging adjacent samples to match the image acquisition rate. Temporal synchronization across the AP, STA, and distributed cameras was achieved via Network Time Protocol (NTP)~\cite{mills1991ntp}, enabling accurate timestamp alignment between image and RSSI samples, forming synchronized image-RSSI pairs for training. The preprocessing parameters were tuned to maintain a Pearson correlation coefficient of approximately 0.90 to 0.95 between the downsampled and preprocessed RSSI, ensuring that the underlying signal trend is preserved while effectively removing noise. Note that the appropriate degree of smoothing may vary by application; more aggressive smoothing may improve model performance but can obscure rapid signal variations critical for latency-sensitive applications.

\begin{table*}[t]
\centering
\resizebox{\textwidth}{!}{%
\begin{tabular}{llcccccccccc}
\toprule
\multirow{3}{*}{\textbf{Configuration}} &
\multirow{3}{*}{\textbf{Model}} &
\multirow{3}{*}{\makecell{\textbf{FLOPs} \\ \textbf{(G)}}} &
\multirow{3}{*}{\makecell{\textbf{Params} \\ \textbf{(M)}}} &
\multicolumn{4}{c}{\textbf{Scene 1}} &
\multicolumn{4}{c}{\textbf{Scene 2}} \\
\cmidrule(r){5-8} \cmidrule(l){9-12}
& & & &
\textbf{RMSE} & \textbf{MAE} & \multirow{2}{*}{\textbf{$r$}} & \multirow{2}{*}{\textbf{$R^{2}$}} &
\textbf{RMSE} & \textbf{MAE} & \multirow{2}{*}{\textbf{$r$}} & \multirow{2}{*}{\textbf{$R^{2}$}} \\
& & & &
\textbf{(dB)} & \textbf{(dB)} & & &
\textbf{(dB)} & \textbf{(dB)} & & \\
\midrule
\multirow{4}{*}{Single-View}
  & SinViT-D (Cam1) & \multirow{2}{*}{1.26} & \multirow{2}{*}{1.46} & 3.2204 & 2.5533 & 0.8300 & 0.6832 & 3.6538 & 2.7849 & 0.7277 & 0.5143 \\
\cmidrule{5-12}
  & SinViT-D (Cam2) & & & 4.1582 & 3.1716 & 0.7009 & 0.4718 & 3.5066 & 2.7455 & 0.7544 & 0.5526 \\
\cmidrule{2-12}
  & SinViT-W (Cam1) & \multirow{2}{*}{2.10} & \multirow{2}{*}{2.90} & 2.9785 & 2.3103 & 0.8562 & 0.7290 & 3.2693 & 2.4849 & 0.7885 & 0.6111 \\
\cmidrule{5-12}
  & SinViT-W (Cam2) & & & 3.8402 & 2.8782 & 0.7457 & 0.5495 & 3.0280 & 2.3853 & 0.8227 & 0.6664 \\
\midrule
\multirow{2}{*}{Multi-View}
  & MulViT-TF    & 1.76 & 1.72 & 2.1953 & 1.6604 & 0.9239 & 0.8528 & 2.5242 & 1.8928 & 0.8770 & 0.7683 \\
\cmidrule{2-12}
  & MulViT-TWDNN & 1.66 & 1.87 & 2.6278 & 2.0549 & 0.8886 & 0.7891 & 2.7238 & 2.0830 & 0.8559 & 0.7302 \\
\bottomrule
\end{tabular}}
\caption{Performance Comparison of Single-View and Multi-View Models for RSSI Estimation Across Various Scenes.}
\label{tab:results}
\end{table*}

\begin{figure*}[t]
    \centering
    \begin{subfigure}[t]{0.49\textwidth}
        \includegraphics[width=\linewidth]{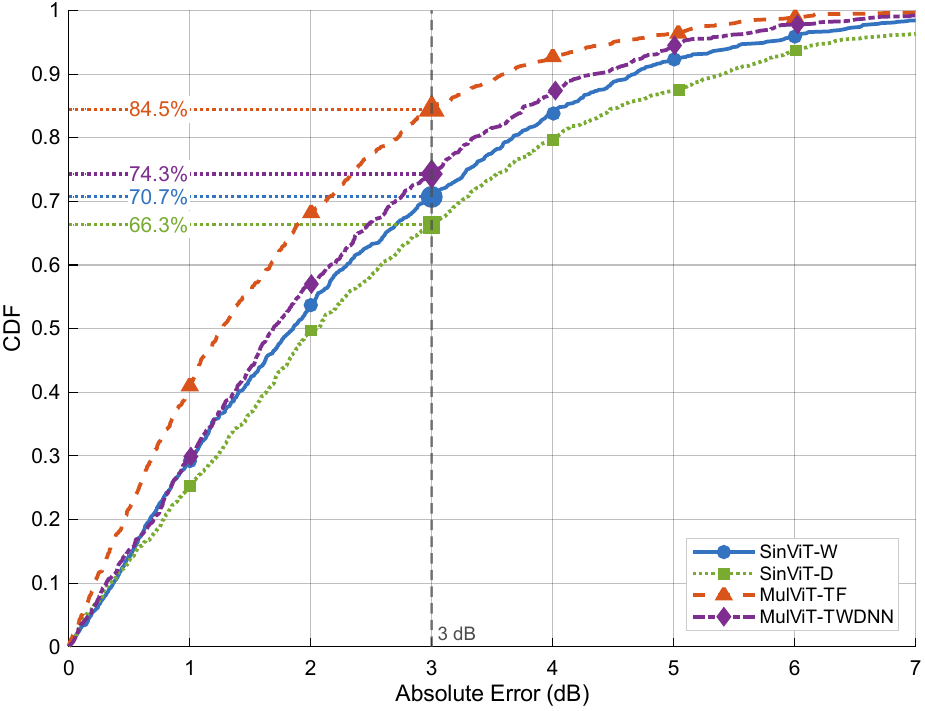}
        \caption{Scene 1}
    \end{subfigure}
    \hfill
    \begin{subfigure}[t]{0.49\textwidth}
        \includegraphics[width=\linewidth]{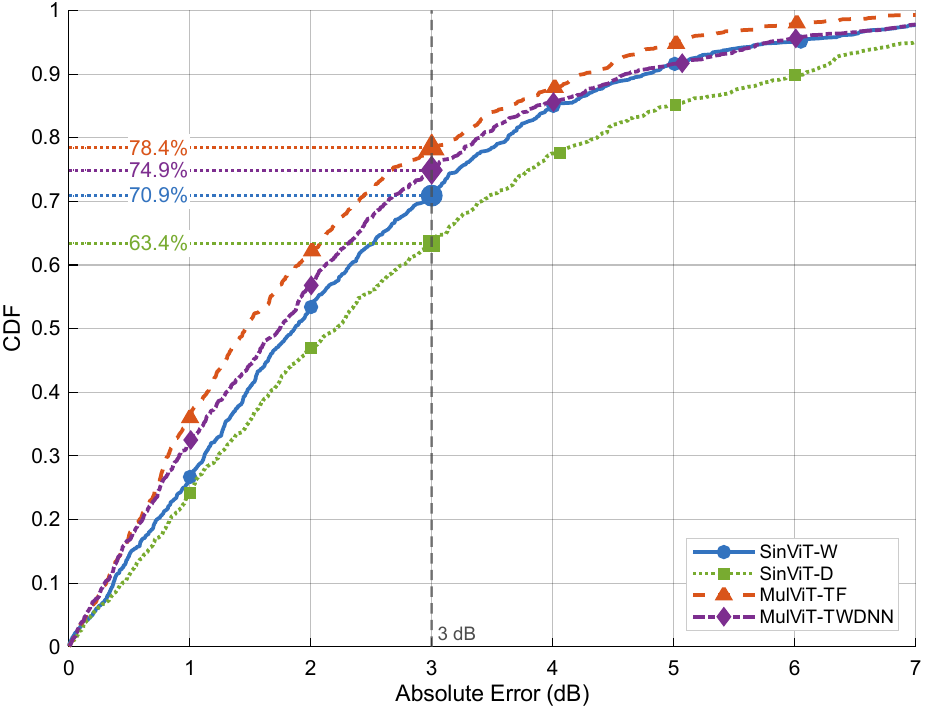}
        \caption{Scene 2}
    \end{subfigure}
    \caption{Empirical CDF of absolute estimation error across models.}
    \label{fig:cdf}
\end{figure*}

The dataset consists of 10,000 samples for each scene, split into 80\%/10\%/10\% for training, validation, and testing, respectively. Input images of resolution $320\times240$ are divided into $16\times16$ patches, yielding 300 patch tokens per image. For single-view models, SinViT-D uses embedding dimension $D{=}96$ with $L{=}12$ Transformer blocks and 3 attention heads, while SinViT-W uses $D{=}192$ with $L{=}6$ blocks and 3 heads. The multi-view models share a common backbone with $D{=}96$, $L{=}6$, and 3 heads per encoder, with two independent encoders. MulViT-TF employs a fusion Transformer block with $L'{=}2$ stacked layers, while MulViT-TWDNN uses 4 residual blocks, each containing two hidden layers of dimension $192$. The MLP head, shared across all models, consists of a single hidden layer of $128$ units with GELU activation. The RSSI labels are normalized using z-score standardization computed from the training set. The models are fine-tuned for 80 and 120 epochs in Phase 1 and 2, respectively, using AdamW~\cite{loshchilov2019adamw} with weight decay $0.1$, learning rate $10^{-4}$, backbone learning rate scale $0.1$, dropout $0.1$, and batch size $32$.

The estimation performance is evaluated using the following metrics: Root Mean Squared Error (RMSE) and Mean Absolute Error (MAE) measure the average magnitude of estimation error in dB, where RMSE is more sensitive to outliers due to the squared term. Pearson correlation coefficient (r) quantifies the linear correlation between estimated and ground-truth RSSI, reflecting how well the model captures the relative signal trend rather than absolute accuracy, and $R^2$ reflects how well the model explains the variance in the ground-truth RSSI. Table~\ref{tab:results} summarizes the performance of all models across both scenes. Among single-view models, SinViT-W consistently outperforms SinViT-D, suggesting that wider embedding is more effective than deeper architecture, though at a higher computational cost due to the quadratic scaling of attention operations with embedding dimension. However, single-view models exhibit significant performance variation between cameras, indicating sensitivity to viewpoint. In contrast, the multi-view models achieve superior and more consistent performance across both scenes compared to all single-view baselines, demonstrating the effectiveness of distributed multi-view fusion. Notably, MulViT-TF attains the best overall results, achieving an RMSE of $2.19$dB, MAE of $1.66$dB, $r$ of $0.92$, and $R^2$ of $0.85$ on Scene 1, and an RMSE of $2.52$dB, MAE of $1.89$dB, $r$ of $0.87$, and $R^2$ of $0.76$ on Scene 2, while requiring lower computational cost than the best-performing single-view baseline, SinViT-W. MulViT-TWDNN also improves over single-view models but falls short of MulViT-TF, demonstrating the advantage of Transformer-based fusion over token-wise feed-forward processing for capturing cross-camera spatial correlations. Fig.~\ref{fig:cdf} further illustrates the empirical Cumulative Distribution Function (CDF) of absolute estimation error across models. We adopt 3dB as the error threshold, which corresponds to a factor-of-two deviation in linear power. MulViT-TF achieves the highest proportion of estimates within this margin, reaching 84.5\% and 78.4\% on Scene 1 and Scene 2, respectively, outperforming all single-view baselines and MulViT-TWDNN. This confirms that distributed multi-view fusion not only reduces average error but also provides more reliable estimates within a practically relevant error bound, and demonstrates the advantage of Transformer-based fusion over token-wise feed-forward processing.

\section{Conclusion}
\label{sec:conclusion}
In this letter, we proposed MulViT-TF, a vision-only distributed multi-view framework for indoor RSSI estimation using spatially distributed RGB cameras. By employing independent ViT encoders per camera and a Transformer-based cross-camera fusion module, the framework effectively captures complementary spatial information under limited FoV coverage and NLoS conditions. Experimental results collected from two distinct real indoor environments demonstrate that MulViT-TF consistently outperforms single-view and token-wise DNN fusion baselines across all evaluation metrics. Future work includes extending the framework to multi-user scenarios and reducing computational overhead as the number of cameras scales. Beyond RSSI, the vision-based approach can be generalized to fundamental wireless tasks such as channel estimation, thereby enabling downstream applications including beamforming, proactive handover, and transmit energy efficiency optimization.



\end{document}